 \documentclass[remotesensing,article,submit,moreauthors,pdftex]{Definitions/mdpi}

\firstpage{1} 
\pubvolume{xx}
\issuenum{1}
\articlenumber{5}
\pubyear{2019}
\copyrightyear{2019}
\preto{\abstractkeywords}{\nolinenumbers} 
\usepackage{subfigure}
\Title{Feature Selection based on Principal Component Analysis for Underwater Source Localization by Deep Learning}

\Author{Xiaoyu Zhu *\orcidA{}, Hefeng Dong , Pierluigi Salvo Rossi\orcidB{} and Martin Landrø\orcidC{}}

\AuthorNames{Xiaoyu Zhu, Hefeng Dong, Pierluigi Salvo Rossi and Martin Landrø}

\address{%
 \quad Department of Electronic Systems,  Norwegian University of Science and Technology, Trondheim, 7491,
Norway; xiaoyu.zhu@ntnu.no; hefeng.dong@ntnu.no; pierluigi.salvorossi@ntnu.no; martin.landro@ntnu.no}

\corres{Correspondence: xiaoyu.zhu@ntnu.no}

\abstract{In this paper, we propose an interpretable feature selection method based on principal component analysis (PCA) and principal component regression (PCR), which can extract important features for underwater source localization by only introducing the source location without other prior information. This feature selection method is combined with a two-step framework for underwater source localization based on the semi-supervised learning scheme. In the framework, the first step utilizes a convolutional autoencoder to extract the latent features from the whole available dataset. The second step performs source localization via an encoder multi-layer perceptron (MLP) trained on a limited labeled portion of the dataset. The proposed approach has been validated on the public dataset SwllEx-96 Event S5. The result shows the framework has appealing accuracy and robustness on the unseen data, especially when the number of data used to train gradually decreases. After feature selection, not only the training stage has a 95\% acceleration but the performance of the framework becomes more robust on the depth and more accurate when the number of labeled data used to train is extremely limited.}

\keyword{Underwater source localization; feature selection; principal component analysis; principal component regression; semi-supervised learning; deep neural network}

\begin{document}
\section{Introduction}
Underwater source localization is a relevant and challenging task in underwater acoustics. The most popular method for source localization is Matched-Field Processing (MFP) \cite{baggeroer1988matched}, which has inspired several works \cite{bogart1992comparative,baggeroer1993overview,mantzel2012compressive,yang2014data}. One of the major drawbacks of the MFP method is the need to compute many “replica” acoustic fields with different environmental parameters via numerical simulations based on the acoustic propagation model. Accuracy of the results is heavily affected by the amount of prior information about the marine environment (e.g. sound speed profile, geoacustic parameters, etc.), which unfortunately is often hard to acquire in real scenarios. 

Artificial intelligence (AI), and primarily data-driven approaches based on machine learning (ML), has become pervasive in many research fields \cite{goodfellow2016deep,chen2020machine}. ML-techniques are commonly divided into supervised and unsupervised learning. The former approach relies on the availability of labeled datasets, i.e. when measurements are paired with ground truth information. The latter refers to the case when unlabeled data are available \cite{ghahramani2003unsupervised}.

Recently, there have been several studies on underwater source localization based on ML using the supervised learning scheme \cite{lefort2017direct,niu2017source,niu2017ship,wang2018underwater,huang2018source,liu2019source,niu2019deep,wang2019deep,lin2020passive}. The general approach of underwater source localization by supervised learning scheme is through the use of acoustic propagation simulation models to create a huge simulation dataset for covering the real scenario. This approach has two main limitations: firstly, creating such a huge simulation dataset is time consuming and requires large computer storage resources; secondly, the set of environmental parameters to create a simulation dataset may not be able to account and adapt for environmental changes in a real-world scenario. 
The latter aspect requires a new simulation process, which may often be unrealistic. 

Apparently, Data-driven ML approaches rely on information extracted from available data, then the need of being able to exploit both labeled and unlabeled data is crucial in many applications, including underwater source localization. 
Semi-supervised learning has been proposed to face this issue in computer vision \cite{zhai2019s4l} and room acoustics \cite{bianco2020semi,hu2020unsupervised}.

Deep learning is famous for its brilliant performance for many tasks, however, the huge computation is the price. In the study of Niu et al. \cite{niu2019deep}, the training time was six days for their ResNet50-1 model and three days for each of the ResNet50-2-x-D models. Each ResNet50-2-x-R model took 15 days to train. 

In the real scenario, the speed of training is vital for real-time localization. To accelerate the training speed, some feature selection methods have been applied in underwater acoustics\cite{zeng2013feature,ouelha2013extension,yang2016underwater,erkmen2008improving}. Feature selection is aiming to find the optimal feature subspace that can express the systematic structure of the raw dataset \cite{zeng2013feature}. Principal component analysis (PCA) is a well-known method of feature selection which can maximize the variance in each principal direction and remove the correlations among the features of the raw dataset  \cite{jackson2005user,zeng2013feature}. Furthermore, the latent relationship between features can be interpreted by studying the loading plot of PCA \cite{westad2003variable}.

In our study, an interpretable feature selection method for underwater source localization based on PCA is proposed. To make the situation closer to the real scenario, a two-step semi-supervised framework, and the data collected by a single hydrophone is used to build and train the neural network, respectively. 

The raw data is firstly preprocessed by discrete Fourier transform and then analyzed by PCA. To select the important features for source localization, a PCA based regression (PCR) is conducted. Based on the absolute value of the regression coefficients of PCR, the important features are selected. Finally, the selected features are fed into the two-step semi-supervised framework for source localization. The framework's structure is built on the encoder of a convolutional autoencoder which is trained in unsupervised-learning mode, and a 4-layer multi-layer perceptron (MLP) which is trained in supervised-learning mode.

The performance of our approach is assessed on the public dataset SWellEx-96 Experiment \cite{murray1996swellex}. 
 
More specifically, the contributions of our work are:
\begin{itemize}
    \item An interpretable approach of feature selection for underwater acoustic source localization is proposed. This approach can reveal the important features related to sources by only introducing the source location without other prior information.
    \item By using the selected features, the training time of the neural networks is significantly decreased with a slight loss of the performance of localization. 
    \item A semi-supervised two-step framework is used for underwater source localization exploiting both unlabeled and labeled data. The performance of the framework is assessed showing appealing behavior in terms of good performance combined with simple implementation and large flexibility.
\end{itemize}

The paper is organized as follows:
Sec.~\ref{sec2} describes the theories of PCA and PCR, as well as the method of feature selection; Sec.~\ref{sec3} presents the two-step framework for underwater source localization; The SwellEx-96 experiment, the data preprocessing, and the schemes of building the training dataset are given in Sec.~\ref{sec4}; In Sec.~\ref{sec5}, a comprehensive analysis of the localization performance between our framework based on the feature selection method and the control groups are described; The selected features are interpreted from both physical and data-science perspective in Sec.~\ref{sec6}. Finally, the conclusion is given in Sec.~\ref{sec7}.
 
%%%%%%%%%%%%%%%%%%%%%%%%%%%%%%%%%%%%%%%%%%
\section{The interpretable feature selection method based on PCA}
\label{sec2}
\subsection{Theory of PCA \citep{esbensen2002multivariate}}
\label{sec2.PCA}

PCA refers to the following decomposition of a column-mean-centered data matrix $\boldsymbol{X}$ of size $N\times K$, where $N$ and $K$ represent the number of samples and the number of features, respectively,
\begin{equation}
\label{eq.1}
     \boldsymbol{X} = \boldsymbol{TP^T}+\boldsymbol{E}
\end{equation}

where $\boldsymbol{T}$ is a score matrix of size $N\times A$ related to the projections of the matrix $\boldsymbol{X}$ into an A-dimensional space, $\boldsymbol{P}$ is a loading matrix of size $K\times A$ related to the projections of the features into the A-dimensional space (with $\boldsymbol{P^TP}=\boldsymbol{I}$), and $\boldsymbol{E}$ is a residual matrix of size $N\times K$.

More specifically, the A-dimensional space is identified via the SVD of $\boldsymbol{X}$ by selecting the first A principal components.

Denoting $\boldsymbol{X}=\boldsymbol{USV^{T}}$ the SVD of $\boldsymbol{X}$ and $\boldsymbol{\hat{U}}$, $\boldsymbol{\hat{S}}$, and $\boldsymbol{\hat{V}}$ the matrices containing the first A columns of $\boldsymbol{U}$, $\boldsymbol{S}$, and $\boldsymbol{V}$, respectively, then we have
\begin{equation}
\label{eq.svd2}
    \begin{aligned}
      \boldsymbol{T}=\boldsymbol{\hat{U}\hat{S}}\\
      \boldsymbol{P}=\boldsymbol{\hat{V}}  
    \end{aligned}
\end{equation}

and $\boldsymbol{\hat{X}}=\boldsymbol{TP^T}$ is called the reconstructed data matrix.

\subsection{Theory of PCR}
The multiple linear regression (MLR) method is given by
\begin{equation}
    \boldsymbol{y}=\boldsymbol{X\theta}+\boldsymbol{e}
    \label{eq.mlr}
\end{equation}
where $\boldsymbol{y}$ is the regression target (in this paper is source location) of size $N\times1$ containing N samples; $\boldsymbol{X}$ is the data matrix as mentioned above; $\boldsymbol{\theta}$ is the regression coefficients of size $K\times1$; and $\boldsymbol{e}$ is the unexplained residuals of $\boldsymbol{y}$. Using ordinary least squares regression \citep{hoy2002combining}, the regression coefficients $\boldsymbol{\hat{\theta}}^{\text{MLR}}$ of size $K\times1$ can be estimated as 
\begin{equation}
    \boldsymbol{\hat{\theta}}^{\text{MLR}}=(\boldsymbol{X}^{T}\boldsymbol{X})^{-1}\boldsymbol{X}^{T}\boldsymbol{y}
    \label{eq.mlr_coef.}
\end{equation}

PCR is the MLR based on the first A PCs extracted from the original data matrix $\boldsymbol{X}$. To estimate the regression parameters $\boldsymbol{\hat{\theta}}^{\text{PCR}}$ of size $A\times1$, the score matrix $\boldsymbol{T}$ is used instead of $\boldsymbol{X}$ in Eq.~\ref{eq.mlr_coef.}:
\begin{equation}
    \boldsymbol{\hat{\theta}}^{\text{PCR}}=(\boldsymbol{T}^{T}\boldsymbol{T})^{-1}\boldsymbol{T}^{T}\boldsymbol{y}
    \label{eq.pcr1}
\end{equation}

\subsection{Method of feature selection}
\label{sec2.FS-method}
The aim of feature selection is to select a set of important variables for accelerating the speed of underwater acoustic source localization. Furthermore, PCA and PCR are highly interpretable methods, the correlation between variables and the significant variables for regression can be revealed by investigating the plot of the loading and the value of regression coefficients, respectively \citep{esbensen2002multivariate}.  

The method of feature selection has 5 steps:
\begin{enumerate}
    \item Conducting mean-centered operation for each column in the data matrix $\boldsymbol{X}$. 
    \item Conducting SVD on the column-mean-centerd data matrix $\boldsymbol{X}$ to calculate the first A PCs ($A=3$ in this paper) as well as build the matrices of the score $\boldsymbol{T}$ and the loading $\boldsymbol{P}$ following Eq \ref{eq.svd2}.
    \item Calculating the regression coefficients $\boldsymbol{\bar{\theta}}$ of size $K\times1$ for each original variables by
    \begin{equation}
        \boldsymbol{\bar{\theta}} = \boldsymbol{P} \boldsymbol{\hat{\theta}}^{\text{PCR}}
    \end{equation}
    \item Ranking the elements in $\boldsymbol{\bar{\theta}}$ with absolute value from high to low. And setting a threshold $\epsilon$ ($\epsilon=0.02$ in this paper) to choose the variables greater than the threshold as the selected features.
    \item A new data matrix $\boldsymbol{\bar X}$ of size $N\times M$ is constructed based on the selected M features.
\end{enumerate}

%%%%%%%%%%%%%%%%%%%%%%%%%%%%%%%%%%%%%%%%%%
\section{The two-step framework for underwater source localization}
\label{sec3}

In the following, we assume that a large-size dataset is available with most of the data being unlabeled and only a small fraction labeled.

\begin{figure}
  \centering
  \subfigure[]{\includegraphics[width=.48\linewidth]{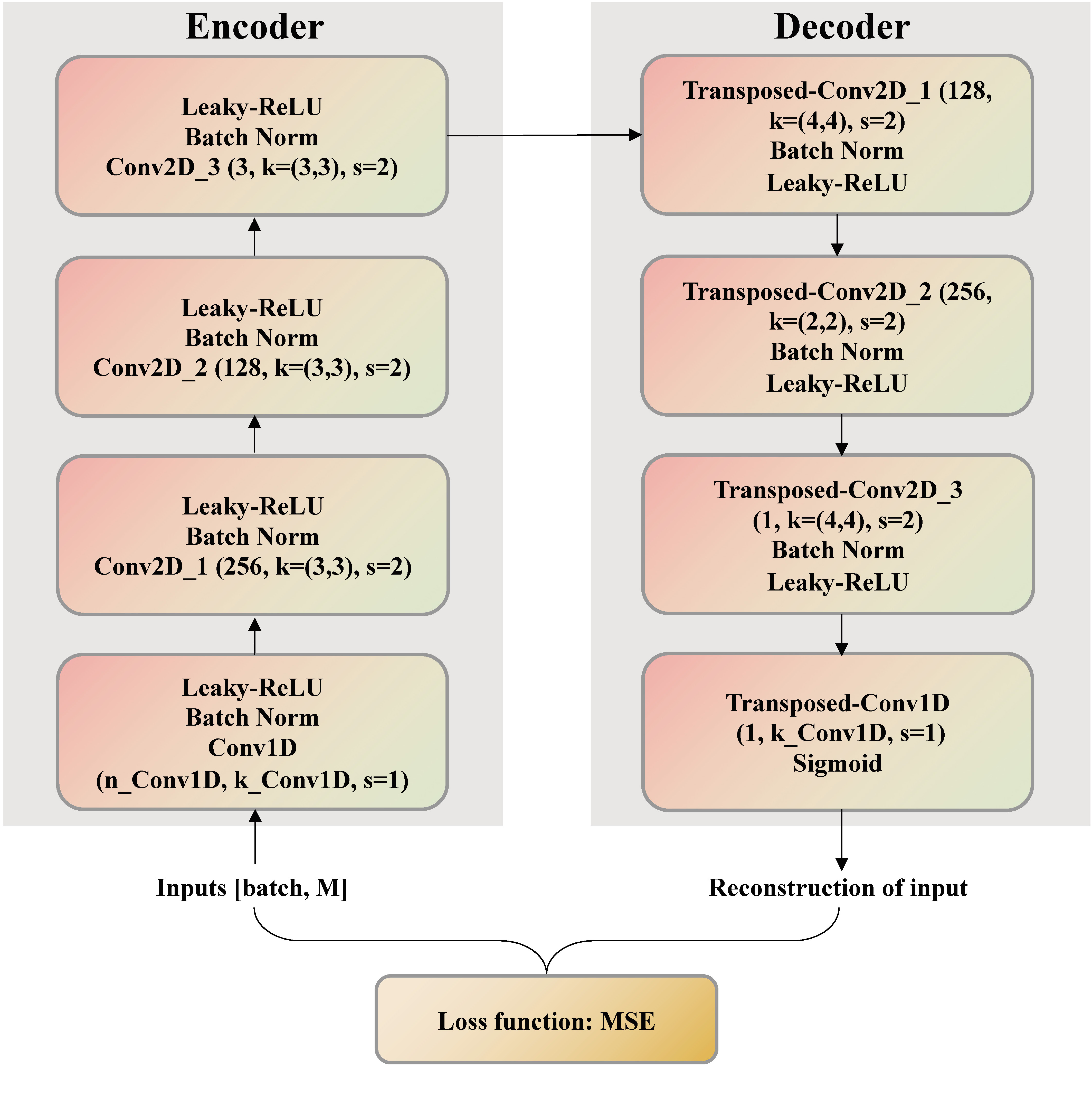}\label{fig.cae}}
  \subfigure[]{\includegraphics[width=.48\linewidth]{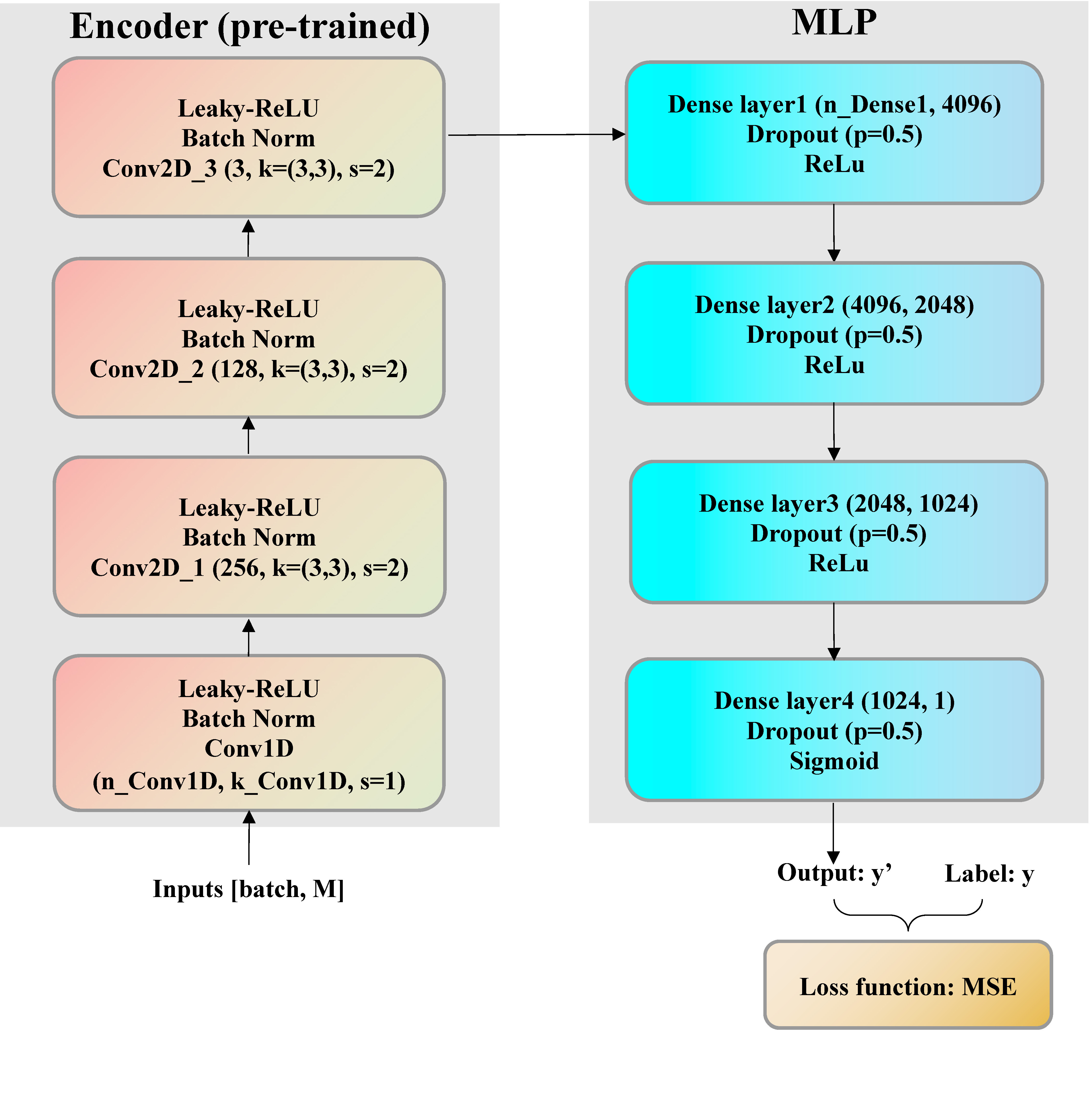}\label{fig.mlp}}
  \caption{Design of our framework: (a) the convolutional autoencoder; (b) the Encoder-MLP localizer.}
  \label{fig1}
\end{figure}
\subsection{Step-one: Train a convolutional autoencoder}
The first step of the framework is to train a convolutional autoencoder (CAE) \cite{hinton1994autoencoders,chen2017deep}. 
The structure of CAE is shown in Fig.~\ref{fig.cae}, where the network consists of an encoder and a decoder. 
The arrows indicate the direction of the data stream. 

The encoder, made of 4 blocks, is used to extract the compressed features from the input data. 
Each block contains a convolution layer (for extracting features), a batch norm layer (for speeding up training), and a leaky-ReLu layer (for operating a nonlinear transform on the data stream). Additionally, the decoder has a dual-symmetric structure as the encoder, and is used to create the reconstruction of the input data from the compressed features. 
After creating the reconstructed input, the Mean Squared Error (MSE) is the selected loss function to measure the difference between the input and the reconstruction. 

It is worth noticing that, in this step, the whole dataset (both the labeled and unlabeled portions) is used to train the network, but only the purely acoustic signals are involved as described later in the paper.
\textbf{The reason to create the training set in this approach is to sufficiently cover all the information in the dataset, which is hard for the purely supervised learning scheme to cover.} 

\subsection{Step-two: Train the Encoder-MLP localizer based on the semi-supervised learning scheme.}

After training the CAE, the second step requires training the Encoder-MLP for localization based on the semi-supervised learning scheme. 
The structure of this model is shown in Fig.~\ref{fig.mlp}, which consists of a pre-trained encoder extracting the compressed features from input data and a 4-layer-MLP estimating the location of the acoustic source based on the compressed features. 
The MLP consists of four blocks, with each block containing a dense layer followed by a dropout layer (for regularization) and a non-linear transform function. 
The sigmoid function is an appropriate choice for the non-linear transform since, during the data preprocessing stage, the regression target, i.e. the horizontal distance between source and receiver, is scaled into the interval $(0,1)$.

Similarly, the arrows in Fig.~\ref{fig.mlp} indicate the direction of the data stream. 
Since the encoder has been trained, its parameters will be frozen during the training stage of the second step. 
After the encoder, the compressed features are fed in the MLP, which will provide the estimated source location as output. 
Finally, the same loss function, i.e.  MSE, is used since the localization task is a regression problem.

\section{Dataset and Preprocessing}
\label{sec4}
\subsection{The SwellEx-96 Experiment}
Vertical linear array (VLA) data from event S5 of the SwellEx-96 experiment are used to illustrate the localization performance of our framework. The event S5 was conducted near San Diego, CA, where the acoustic source started its track of all arrays and proceeded northward at a speed of 5 knots (2.5 m/s). The source had two sub-sources, a shallow one was at a depth of 9 m and a deep one at 54 m. The sampling rate of the data was 1500 Hz and the recording time of the data was 75 min. The VLA contained 21 sensors equally spaced between 94.125 m and 212.25 m. The water depth was 216.5 m. Additionally, the horizontal distance between the source and VLA is also provided in the dataset. More detailed information of this event can be found in \cite{murray1996swellex}. 

\subsection{Preprocessing and feature selection}

In this paper, the underwater acoustic signals collected by a single receiver are transformed into the frequency domain. 
We calculate the spectrum without overlap for every 1 s slices of the signal and arrange it in a matrix (namely \textbf{features}) $\boldsymbol{X}$ format with the shape of $4500\times 750$, where each row is related to one slice. 
More specifically, 4500 is the total number of time-step ($75 min = 4500 sec$) and 750 is the number of frequencies. 
In the matrix, each row corresponds to one single time-step, and each column corresponds to one single frequency. 

Besides the acoustic signals, the horizontal distance between the source and VLA was provided in the original dataset, which can be expressed as a vector $\boldsymbol{y}$ (namely \textbf{labels}) with the shape of $4500\times1$, where $4500$ indicates the total number of time-step and $1$ indicates the distance at each time-step.

For the training stability of our framework, the features $\boldsymbol{X}$ and labels $\boldsymbol{y}$ are scaled into interval $(0,1)$ by the Min-Max Scaling method:
\begin{equation}
\begin{aligned}
\boldsymbol{ X} = \frac{\boldsymbol{X}-\boldsymbol{X}_{\text{min}}}{\boldsymbol{X}_{\text{max}}-\boldsymbol{X}_{\text{min}}} \;,\qquad
\boldsymbol{ y} = \frac{\boldsymbol{y}-\boldsymbol{y}_{\text{min}}}{\boldsymbol{y}_{\text{max}}-\boldsymbol{y}_{\text{min}}}  \;.
\end{aligned}
\label{eq.minmax}
\end{equation}

After preprocessing, the feature selection is conducted followed the steps in Sec.~\ref{sec2} based on $\boldsymbol{X}$ and $\boldsymbol{y}$. 
\subsection{Schemes for building the dataset}
\label{sec.4.3}
For step-one, the dataset for CAE is expressed as:
\begin{equation}
    [\boldsymbol{\bar X}]=[\boldsymbol{\bar x}_i]_{i=1}^{N}
\end{equation}
where $\boldsymbol{\bar X} $ is the features in matrix form, $\boldsymbol{\bar x}_i$ is a row-vector with the length of M (the number of selected features), corresponding to the $i$th row of the features matrix $\boldsymbol{\bar X}$ and $N$ is the number of time-step. 

For step-two, the dataset for Encoder-MLP localizer is expressed as:
\begin{equation}
    [\boldsymbol{\bar X}, \boldsymbol{y}]=[\boldsymbol{\bar x}_i,  y_i]_{i=1}^{N}
\end{equation}
where $\boldsymbol{\bar X} $ and $\boldsymbol{y}$ is the features and labels in matrix form. And $y_i$ is the $i$th element in labels vector $\boldsymbol{y}$.

\subsection{Scheme of separating training and test dataset}
\label{sec.4.4}
To illustrate the performance of the semi-supervised framework as the number of labeled datasets decreases, 50\%, 25\%, and 12.5\% of the whole labeled dataset are chosen, respectively, as the training set of step-two. 

Since source localization is a regression task, the labels in the training set of step-two should cover the whole interval of the horizontal distance between the source and receiver. As described above, the total number of time-step is 4500, which can be expressed by the index $i\in (1,4500)$. The schemes of separating training and test dataset for step-two are:

\subsubsection{Using 50\% data to build training set}
\begin{equation}
    \begin{aligned}
        \text{Training set} &:(\boldsymbol{\bar x}_i, y_i) && \forall i: mod(i,2)=1\\
        \text{Test set} &: (\boldsymbol{\bar x}_i, y_i) && \forall i: mod(i,2)\neq 1
    \end{aligned}
\end{equation}

\subsubsection{Using 25\% data to build training set}
\begin{equation}
    \begin{aligned}
        \text{Training set} &:(\boldsymbol{\bar x}_i, y_i) && \forall i: mod(i,4)=1\\
        \text{Test set} &: (\boldsymbol{\bar x}_i, y_i) && \forall i: mod(i,4)\neq 1
    \end{aligned}
\end{equation}

\subsubsection{Using 12.5\% data to build training set}
\begin{equation}
    \begin{aligned}
        \text{Training set} &:(\boldsymbol{\bar x}_i, y_i) && \forall i: mod(i,8)=1\\
        \text{Test set} &: (\boldsymbol{\bar x}_i, y_i) && \forall i: mod(i,8)\neq 1
    \end{aligned}
\end{equation}

To show the influence of different depths, receivers no.1 (top), no.10 (middle), and no.21 (bottom) are chosen to build the dataset, respectively. For each receiver, there are 3 choices of the percentage to build the training dataset. Totally, there are $3\times3=9$ training datasets are built.

\subsection{Control group}
\subsubsection{Control group for the semi-supervised framework}
To make a fair comparison, we trained a neural network with the same structure as the framework by the purely supervised learning scheme. 
\subsubsection{Control group for the feature selection method}
To show the performance of our feature selection method, a framework without feature selection is trained in the same way. 
The matrix containing the whole features $\boldsymbol{X}$ of size $4500\times750$ is calculated from Eq. \ref{eq.minmax} and used to build the dataset following the scheme described in Sec. \ref{sec.4.3} and Sec. \ref{sec.4.4}.

\section{Performance of source localization}
\label{sec5}
\subsection{Hyperparameters of the framework}
In Fig.~\ref{fig1}, the output channel (n
\_Conv1D) and the kernel size (k\_Conv1D) of the 1D-convolutional layer, as well as the input channel (n\_Dense1) of the first dense layer, are not fixed. This is because the size of input features are various between dataset collected by different sensors.

For the training dataset without feature selection,
\begin{itemize}
    \item n\_Conv1D= 738
    \item k\_Conv1D= 13
    \item n\_Dense1= 24843
\end{itemize}

For the training dataset with feature selection, 
\begin{itemize}
    \item n\_Conv1D= 114 
    \item k\_Conv1D= $M-113$
    \item n\_Dense1= 507
\end{itemize}

After feature selection, the number of selected features $M$ is shown in Table. \ref{table.n_feature}.
\begin{table}[H]
\caption{Number of selected features among different receivers.}
\centering
\begin{tabular}{cccc}
\toprule
Receiver&50\%&25\%&12.5\%\\
\midrule
No.1 &121&122&125\\
No.10 &129&137&134\\
No.21 &126&126&127\\
\bottomrule
\end{tabular}
\label{table.n_feature}
\end{table}

To train the framework, the learning rate for step-one and step-two is 1e-4 and 5e-5, respectively. The optimization scheme is Adam and the epoch is 100 for each step.

All the networks mentioned in this paper were trained using one NVIDIA RTX 2080Ti GPU card. 
\subsection{Examining the performance when removing some 2D-convolutional layers of the framework after feature selection}
\label{sec.best_FS}

The function of the encoder is to compress the original dataset and create its compressed expression, which is similar to our manual feature selection method. To find out the best structure for the framework using feature selection, the number of 2D-convolutional layers of the encoder, and the corresponding number of transposed 2D-convolutional layers of the decoder are gradually decreased. After re-training the modified CAE, the step-two is conducted as before. The structure of CAE after removing one and two 2D-convolutional layers are shown in Table \ref{table.remove1} and \ref{table.remove2}, respectively. The performance will be discussed in Sec. \ref{overall_discussion}.
\begin{table}[H]
    \caption{Structure 1: Removing one 2D-convolutional layer of the encoder}
    \centering
    \begin{tabular}{ccccc}
    \toprule
      &Block &Output channel&Kernel size&Stride\\
    \midrule
    Encoder &Conv1D&114&M-113&1\\
             &Conv2D\_1&128&$3\times3$&2\\
             &Conv2D\_2&3&$3\times3$&2\\
    Decoder &Transposed-Conv2D\_1&128&$4\times4$&2\\
            &Transposed-Conv2D\_2&1&$4\times4$&2\\
            &Transposed-Conv1D&1&M-113&1\\
    \bottomrule
    \end{tabular}
    \label{table.remove1}
    \end{table}
\begin{table}[H]
    \caption{Structure 2: Removing two 2D-convolutional layer of the encoder}
    \centering
    \begin{tabular}{ccccc}
    \toprule
      &Block &Output channel&Kernel size&Stride\\
    \midrule
    Encoder &Conv1D&114&M-113&1\\
             &Conv2D\_1&3&$3\times3$&2\\
    
    Decoder &Transposed-Conv2D\_1&1&$4\times4$&2\\
            &Transposed-Conv1D&1&M-113&1\\
    \bottomrule
    \end{tabular}
    \label{table.remove2}
\end{table}

\subsection{Overall analysis of the localization performance}
\label{overall_discussion}
To make a comprehensive comparison, 4 pairs of networks are tested on the data of all receivers and trained separately based on the data collected by receivers no.1, no.10, and no. 21. One pair is trained without feature selection, the rest are all trained with the feature selection method proposed by this paper. For the rest 3 pairs of networks, one has the same number of layers as the networks trained without feature selection; others have the structures shown in Tables \ref{table.remove1} and \ref{table.remove2}, respectively. Additionally, each pair of networks is consists of the framework trained by the semi-supervised learning scheme and the same network of step-two trained by the purely supervised learning scheme.

\subsubsection{Comparison between the framework and the purely supervised learning scheme after feature selection.}

After feature selection and tested on all receivers, performances of our framework and the purely supervised learning scheme are shown in Table \ref{no.1}. In the Table, the first row indicates the percentage of the data used to build the training set. In the first column, R1 to R21 indicates receivers no.1 to no.21, respectively. Additionally, the Mean indicates the average of MSE on all receivers. The bold numbers indicate the lower values of MSE in every pair of our framework and the purely supervised learning scheme, which means the model has a better performance of source localization.

Observing Table \ref{no.1}, interesting phenomena can be found:
\begin{enumerate}
    \item Performance of the purely supervised learning scheme:
    
    The network trained by the supervised learning scheme can attain the lower MSE only when the test set is chosen near the sensor used to build the training set. When the test set is far from the sensor used to train, its performance is getting worse dramatically. This trend is more obvious when the percentage of data used to train decreases. \textbf{This shows the limitation of the purely supervised learning scheme: when the labeled training set is limited, the generalization ability of the model is poor.}
    \item Performance of our framework:
    
    Compared to the purely supervised learning scheme, \textbf{our framework is more robust and has much lower MSE} on the data collected by those sensors which are far from the sensor used to build the training set, even though its performance on the data collected by the sensors near the sensor used to train is a bit poorer. This trend is more obvious when the percentage of data used to train decreases.
    
    \item Comparison of the different percentages used to train:
    
    When the percentage of the data used to build the training set decreases, the performance of both schemes becomes worse. However, \textbf{the degree of performance degradation of our framework is smaller than the purely supervised learning scheme.} 
\end{enumerate}

\begin{table}[ht]
    \caption{The MSE of models with feature selection trained on receiver no.1}
    \centering
    \begin{tabular}{ccccccc}
    \toprule
      &\multicolumn{2}{c}{50\%}&\multicolumn{2}{c}{25\%}&\multicolumn{2}{c}{12.5\%}\\
    
      & Framework & Supervised & Framework & Supervised & Framework & Supervised\\
    \midrule
     R1 & \textbf{0.22} & \textbf{0.22} & \textbf{0.31} & \textbf{0.31} & \textbf{0.40} & 0.44 \\
     R2 & \textbf{0.31} & 0.33 & \textbf{0.36} & 0.39 & \textbf{0.48} & 0.51 \\
     R3 & \textbf{0.34} & 0.37 & \textbf{0.39} & 0.42 & \textbf{0.44} & 0.50 \\
     R4 & \textbf{0.35} & 0.39 & \textbf{0.43} & 0.44 & \textbf{0.45} & 0.55 \\
     R5 & \textbf{0.41} & 0.44 & 0.47 & \textbf{0.44} & \textbf{0.47} & 0.58 \\
     R6 & \textbf{0.39} & 0.42 & \textbf{0.4} & 0.42 & \textbf{0.42} & 0.54 \\
     R7 & \textbf{0.44} & 0.47 & \textbf{0.46} & \textbf{0.46} & \textbf{0.5} & 0.58 \\
     R8 & \textbf{0.38} & 0.4 & \textbf{0.38} & 0.43 & \textbf{0.42} & 0.48 \\
     R9 & \textbf{0.36} & 0.42 & \textbf{0.4} & 0.41 & \textbf{0.4} & 0.54 \\
     R10& \textbf{0.39} & 0.43 & 0.46 & \textbf{0.45} & \textbf{0.48} & 0.59 \\
     R11& \textbf{0.4} & 0.5 & \textbf{0.49} & 0.52 & \textbf{0.49} & 0.64 \\
     R12& \textbf{0.39} & 0.4 & \textbf{0.43} & 0.46 & \textbf{0.46} & 0.54 \\
     R13& \textbf{0.37} & 0.45 & \textbf{0.44} & 0.48 & \textbf{0.5} & 0.67 \\
     R14& \textbf{0.4} & 0.49 & \textbf{0.49} & \textbf{0.49} & \textbf{0.49} & 0.63 \\
     R15& \textbf{0.4} & 0.41 & \textbf{0.4} & 0.43 & \textbf{0.47} & 0.48 \\
     R16& \textbf{0.47} & 0.49 & \textbf{0.48} & 0.5 & \textbf{0.56} & 0.59 \\
     R17& 0.56 & \textbf{0.5} & \textbf{0.51} & 0.57 & 0.62 & \textbf{0.61} \\
     R18& \textbf{0.43} & 0.47 & \textbf{0.45} & 0.51 & \textbf{0.51} & 0.57 \\
     R19& \textbf{0.41} & 0.51 & \textbf{0.5} & 0.52 & \textbf{0.48} & 0.63 \\
     R20& \textbf{0.43} & 0.51 & \textbf{0.51} & 0.53 & \textbf{0.53} & 0.67 \\
     R21& \textbf{0.45} & 0.53 & \textbf{0.53} & 0.59 & \textbf{0.58} & 0.8 \\
     
     Mean& \textbf{0.40} & 0.44 & \textbf{0.44} & 0.47 & \textbf{0.48} & 0.58 \\
    \bottomrule
    \end{tabular}
    \label{no.1}
\end{table}

\subsubsection{Comparison of the mean MSE and training time between the networks with and without feature selection}
The performance between the networks with and without feature selection are shown in Table \ref{table.with_without_feature_selection}. In the table, the residual illustrates the difference of the mean MSE between networks, and the percentage of the residual illustrates the performance improvement (positive value) and degradation (negative value) by feature selection. They are calculated by
\begin{equation}
\begin{aligned}
    &Residual=MSE_{~Without~FS}-MSE_{~With~FS}\\
    &Percentage~of~Residual=\frac{Residual}{MSE_{~Without~FS}}
\end{aligned}
\end{equation}

Observing Table \ref{table.with_without_feature_selection}, phenomena can be found:

    \begin{enumerate}
    \item When the percentage of data used to train is 50\% and 25\%, the performance of the framework trained on R1 and R10 has some degradation (12.82\% to 17.65\%).  However, when the percentage of data used to train is 12.5\%, the performance of the framework trained on R1 and R10 has a slight improvement (7.69\%)  and degradation (4\%), respectively.
    \item Trained on R21, the framework's performance has significant improvements, which are 17.24\%, 31.82\%, and 44.71\% when the percentage of data used to train is 50\%, 25\%, and 12.5\%, respectively.
    \item Compared to the framework, the performance of the purely supervised learning scheme gains more improvement (14.04\% to 50.47\%) after feature selection. The performance degradation only happens when it is trained on 50\% R1, 50\% R10, and 25\% R10.
\end{enumerate}

The training time of the networks are shown in Table \ref{table.training_time_with_without_feature_selection}. This table illustrates that \textbf{the training time is reduced significantly after feature selection for both framework and the purely supervised learning scheme}. 
\begin{table}[H]
\tablesize{\footnotesize}
    \caption{Comparison of mean MSE for networks with and without feature selection (FS)}
    \centering
    \begin{tabular}{cccccccc}
    \toprule
     & & \multicolumn{2}{c}{Trained on R1}&\multicolumn{2}{c}{Trained on R10}&\multicolumn{2}{c}{Trained on R21}\\
     & & Framework & Supervised & Framework & Supervised & Framework & Supervised\\
     \hline
     50\%&Without FS& 0.34 &0.43 &0.35 &0.43 &0.58& 0.57\\
         &With FS&0.40&0.44&0.41&0.48&0.48&0.49\\ 
         &Residual&-0.06&-0.01&-0.06&-0.05&0.10&0.08\\
         &Percentage of residual&-17.65\%&-0.02\%&-17.14\%&-11.63\%&17.24\%&14.04\%\\
     \hline
     25\%&Without FS& 0.39  &0.57  &0.40  &0.52  &0.66 & 0.79 \\
         &With FS&0.44 &0.47 &0.46 &0.55 &0.45 &0.53 \\
         &Residual&-0.05&0.10&-0.06&-0.03&0.21&0.26\\
         &Percentage of residual&-12.82\%&17.54\%&-15.00\%&-5.77\%&31.82\%&32.91\%\\
         
     \hline
     12.5\%&Without FS& 0.52&0.78&0.50&0.78&0.85&1.07 \\
         &With FS&0.48&0.58&0.52&0.58&0.47&0.53 \\
         &Residual&0.04&0.20 &-0.02 &0.20 &0.38 &0.54 \\
         &Percentage of residual&7.69\%&25.64\%&-4.00\%&25.64\%&44.71\%&50.47\%\\
    \bottomrule
    \end{tabular}
    \label{table.with_without_feature_selection}
\end{table}
\begin{table}[H]
\tablesize{\footnotesize}
    \caption{Comparison of the training time for networks with and without FS}
    \centering
    \begin{tabular}{cccccccc}
    \toprule
     & & \multicolumn{2}{c}{50\%}&\multicolumn{2}{c}{25\%}&\multicolumn{2}{c}{12.5\%}\\
     & & Framework & Supervised & Framework & Supervised & Framework & Supervised\\
     \hline
     Stet-one&Without FS& 3h 30m 45s &- &3h 30m 45s &- &3h 30 m45s &-\\
            & With FS&7m 7s&-&6m 59s&-&7m 4s&-\\
            & Percentage of reduction&96.62\%&-&96.68\%&-&96.65\%&-\\
    \hline
    Step-two&Without FS& 1h3m46s& 1h 30m 58s&59m 33s&1h 26m 31s&54m 12s& 1h 8m 28s\\
            & With FS&3m 18s&4m 31s&2m 19s&2m 57s&1m 50s&2m 9s\\
            &Percentage of reduction&94.82\%&95.03\%&96.11\%&96.59\%&96.62\%&96.86\%\\
    \bottomrule
    \end{tabular}
    \label{table.training_time_with_without_feature_selection}
\end{table}

\subsubsection{The best structure for the framework after the feature selection}
As mentioned in Sec. \ref{sec.best_FS}, the comparison of the mean MSE and the training time between different structures of networks are shown in Tables \ref{table.mean_MSE} and \ref{table.training_time}, respectively.

\begin{table}[H]
    \caption{Comparison of mean MSE for different structures of the networks after feature selection}
    \centering
    \begin{tabular}{cccccccc}
    \toprule
     & & \multicolumn{2}{c}{Trained on R1}&\multicolumn{2}{c}{Trained on R10}&\multicolumn{2}{c}{Trained on R21}\\
     & & Framework & Supervised & Framework & Supervised & Framework & Supervised\\
     \hline
     50\%&Original structure&0.40&0.44&0.41&0.48&0.48&0.49\\ 
         &Structure 1&0.41&0.41&0.40&0.45&0.44&0.43\\
         &Structure 2&0.37&0.42&0.41&0.42&0.44&0.45\\ 
     \hline
     25\%&Original structure&0.44 &0.47 &0.46 &0.55 &0.45 &0.53 \\
         &Structure 1&0.42 &0.47 &0.43 &0.54 &0.44 &0.48 \\
         &Structure 2&0.46 &0.43 &0.45 &0.46 &0.44 &0.45 \\
     \hline
     12.5\%&Original structure&0.48&0.58&0.52&0.58&0.47&0.53 \\
         &Structure 1&0.51&0.55&0.50&0.62&0.46&0.56\\
         &Structure 2&0.52&0.49&0.51&0.53&0.46&0.48\\
    \bottomrule
    \end{tabular}
    \label{table.mean_MSE}
\end{table}

\begin{table}[H]
    \caption{Comparison of the training time for different structures of the networks after feature selection}
    \centering
    \begin{tabular}{cccccccc}
    \toprule
     & & \multicolumn{2}{c}{50\%}&\multicolumn{2}{c}{25\%}&\multicolumn{2}{c}{12.5\%}\\
     & & Framework & Supervised & Framework & Supervised & Framework & Supervised\\
     \hline
     Step-one & Original structure&7m 7s&-&6m 59s&-&7m 4s&-\\
              & Structure 1&5m 27s&-&5m 27s&-&5m 17s	-\\
              &Structure 2&3m 42s&-&3m 42s&-&3m 44s&-\\
    \hline  
    Step-two&Original structure&3m 18s&4m 31s&2m 19s&2m 57s&1m 50s&2m 9s\\
            & Structure 1&3m 58s&4m 38s&2m 31s&2m 51s&1m 48s&1m 59s\\
            &Structure 2&8m 14s&8m 44s&4m 47s&5m 1s&3m 3s&3m 12s\\
    \hline
    Total&Original structure&10m 25s&4m 31s&9m 18s&2m 57s&8m 54s&2m 9s\\
            & Structure 1&9m 25s&4m 38s&7m 58s&2m 51s&7m 5s&1m 59s\\
            &Structure 2&11m 56s&8m 44s&8m 29s&5m 1s&6m 47s&3m 12s\\
    \bottomrule
    \end{tabular}
    \label{table.training_time}
\end{table}

According to the Tables, interesting phenomena can be found:
\begin{enumerate}
    \item For the framework, Structure 1 attains the lowest MSE except for trained on 50\% R1 and 12.5\% R1.
    \item For the purely supervised learning scheme, Structure 2 attains the lowest MSE with a slight improvement compared to Structure 1 when the percentages of data used to train are 25\% and 12.5\%.  
    \item Structure 1 shows the best performance for training time reduction. \textbf{Considering both MSE and training time, the best structure after the feature selection is Structure 1.}  

\end{enumerate}
\subsubsection{Conclusion of the performance analysis}
\begin{enumerate}
    \item When the number of labeled data is gradually decreasing, the power of the framework with the semi-supervised learning scheme will be illustrated.
    \item The feature selection method is beneficial for both the framework and the purely supervised learning scheme, which can significantly decrease the training time with a slight loss of the performance of localization. 
    \item After feature selection, the difference in performance between different depths is not significant, which means it can increase the robustness of the receiver depth selection. Especially, it can also improve the performance of the networks trained based on receiver no.21.
\end{enumerate}

\begin{figure}[H]
    \begin{center}
    \includegraphics[width = .5\textwidth]{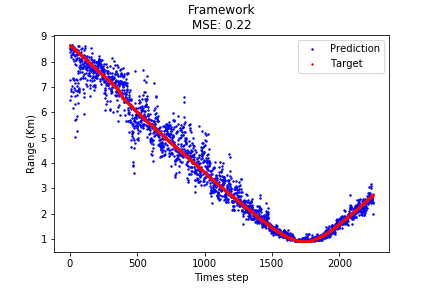}
    \caption{\label{fig.illustration}{Illustration of localization result}}
    \end{center}
\end{figure}

To have an intuitive view of the performance, Fig. \ref{fig.illustration} shows the localization result of our framework trained on 50\% R1 after feature selection.

\section{Discussion of feature selection}
\label{sec6}
\subsection{Details of the sources in the S5 event}
According to the details on the website of the S5 event (http://swellex96.ucsd.edu/s5.htm), the deep source (J-15) transmitted 5 sets of 13 tones between 49Hz and 400Hz. The first set of tones is projected at maximum transmitted levels of 158 dB. The second set of tones are projected with levels of 132 dB. The subsequent sets (3rd, 4th, and 5th) are each projected 4 dB down from the previous set. The shallow source transmitted 9 tones between 109 Hz and 385 Hz.

\subsection{Interpretation of the selected features}

After the feature selection described in Sec. \ref{sec2.FS-method}, the matrix $\boldsymbol{\bar X}$ of size $N\times M$ containing selected features is created. To interpret the selected features, another PCA is conducted on this matrix. To investigate the correlation structure between the features and the PCs, correlation loading is calculated based on the method proposed by Frank Westad et al.~\citep{westad2003variable}.

\begin{figure}[ht]
\begin{center}
\includegraphics[width = \textwidth]{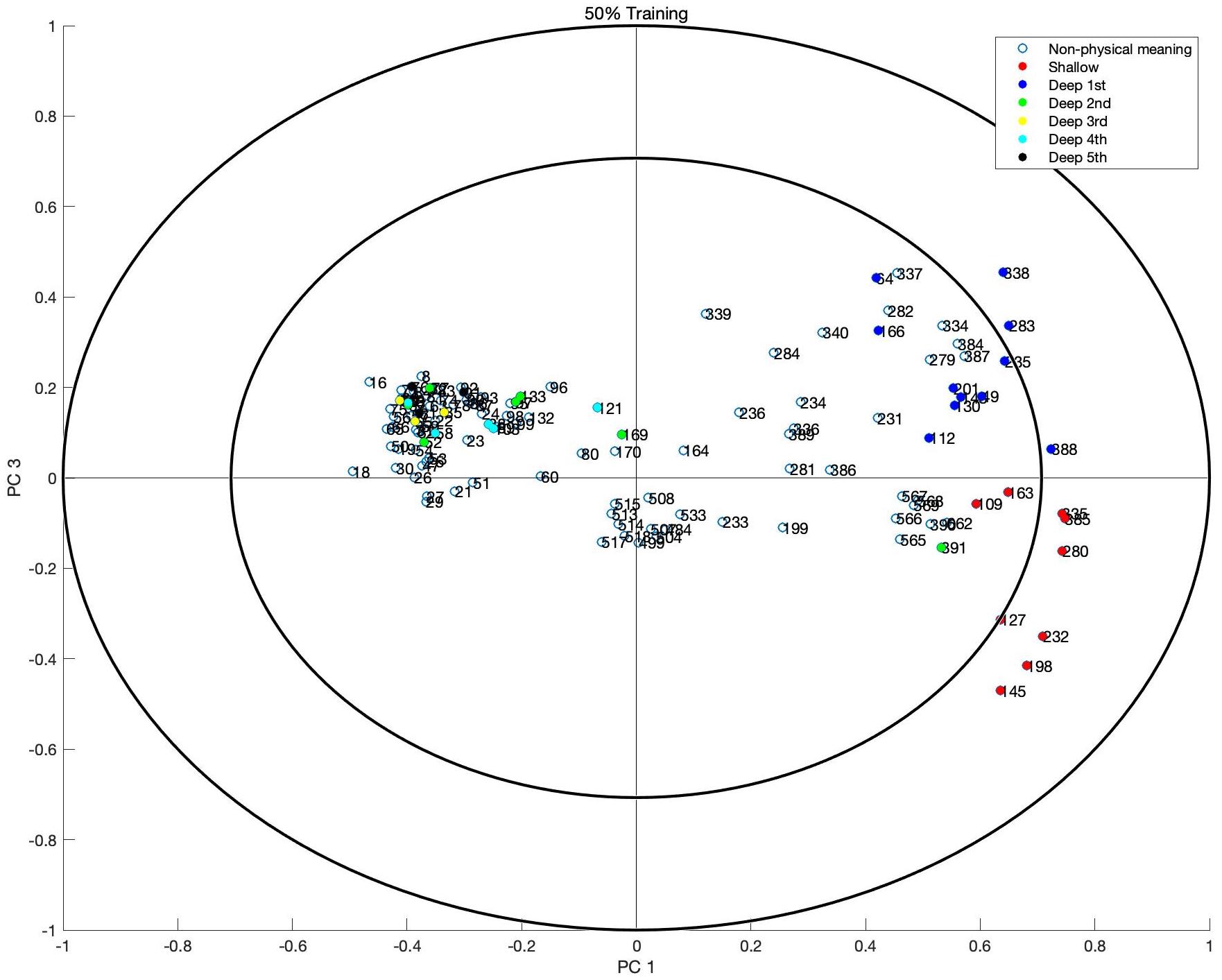}
\caption{\label{fig:interpreate}{Correlation loading plot}}
\end{center}
\end{figure}

The training dataset using 50\% data collected by receiver no.21 was used for interpretation after feature selection. As shown in Fig.~\ref{fig:interpreate}, the abscissa is PC1 and the ordinate is PC3. There are 2 circles in the plot, in which the inner and outer one indicates 50\% and 100\% explained variance, respectively. The points between the two circles are the significant features that can explain at least a 50\% variance of the data. And the legends with different colors illustrate different sets of tones. 

From the correlation loading plot in Fig. \ref{fig:interpreate}, phenomena with physical meanings can be found:

1. Along PC1, the frequencies related to the high transmitted signal level are in the positive half-axis. The frequencies related to lower energy levels are in the negative half-axis. For more details:
    \begin{itemize}
        \item 7 frequencies (127, 145, 198, 232, 280, 335, 385 Hz) of the shallow source and 3 frequencies (238, 338, 388 Hz) of the deep source are in the area between two circles, which means they are significant features.
        \item The rest frequencies of the shallow source and the highest transmitted level (and also the tone with 391Hz related to the second transmitted level) of the deep source are also close to the boundary of the inner circle, which means they still have some importance for the data.
        \item Expect for frequencies of the highest and 391Hz of the second transmitted levels, the rest frequencies are closer to the origin, which means they are less significant and related to the noise from the statistical perspective.
    \end{itemize}
    
2. Along PC3, the frequencies related to the shallow source are in the negative half-axis (except for the tone with 391 Hz); The frequencies related to the deep source are in the positive half-axis.

\textbf{An important point needs to be emphasized that, except for the location of the sources, our feature selection method does not need any prior information.}  The interpretation section is to illustrate the rationality of our feature selection method from a physical perspective. From not only the data science but also the acoustic point of view, the most important features are selected after feature selection. 

\section{Conclusion}
\label{sec7}
In this paper, we propose an interpretable feature selection method based on PCA and PCR, which can extract important features for underwater source localization by only introducing the source location without other prior information. This feature selection method is combined with a two-step framework for underwater source localization based on the semi-supervised learning scheme. The first step utilizes a CAE to extract the latent features for sufficiently using the information of the whole dataset. Then, an Encoder-MLP network, trained on a limited amount of labeled dataset, is used for source localization.

Based on a public dataset, the SwellEx-96 Event S5, the performance of our feature selection method and the two-step framework has been demonstrated. The result shows that the framework is more robust on the unseen data, especially when the number of labeled data used to train gradually decreases. After feature selection, the training time is significantly decreased (averagely 95\%). Especially when the percentage of data used to train is 12.5\%,  the feature selection method can improve the performance of both semi-supervised learning and purely supervised learning. The interpretation of the selected features also demonstrates the rationality of our feature selection method from a physical perspective.

It needs to be mentioned that the structure of the network used in this paper is just a demo for showing the performance of our framework. The more complex and powerful networks can be applied in this framework, based on our anticipation, the performance of source localization will be better as long as the network has been trained appropriately and well.

\authorcontributions{Formal analysis, X.Z., H.D., P.S.R, and M.L.; Funding acquisition, H.D. and M.L.; Methodology, X.Z., H.D., and P.S.R; Resources, X.Z., H.D., and P.S.R; Software X.Z. and H.D.; Writing-original draft, X.Z.; Writing-review and editing, H.D., P.S.R, and M.L. All authors have read and agreed to the published version
of the manuscript.}

%%%%%%%%%%%%%%%%%%%%%%%%%%%%%%%%%%%%%%%%%%
\funding{This research was funded by the Norwegian Research Council and the industry partners of the GAMES consortium at NTNU for financial support (Grant No. 294404).}

%%%%%%%%%%%%%%%%%%%%%%%%%%%%%%%%%%%%%%%%%%
\acknowledgments{X.Z. would like to acknowledge the China Scholarship Council (CSC) for the fellowship support (No. 201903170205).}

%%%%%%%%%%%%%%%%%%%%%%%%%%%%%%%%%%%%%%%%%%
\conflictsofinterest{The authors declare no conflict of interest.} 

%%%%%%%%%%%%%%%%%%%%%%%%%%%%%%%%%%%%%%%%%%
%% optional
\abbreviations{The following abbreviations are used in this manuscript:\\

\noindent 
\begin{tabular}{@{}ll}
PCA & Principal component analysis\\
PCR & Principal component regression\\
MLR & Multiple linear regression\\
PC & Principal component\\
MLP & Multi-layer perceptron\\
MFP & Matched Field Processing\\
AI & Artificial intelligence\\
ML & Machine learning\\
CAE & Convolutional autoencoder\\
MSE & Mean squared error\\
VLA & Vertical linear array\\
FS & Feature selection
\end{tabular}}

%=====================================
% References, variant B: external bibliography
%=====================================
\reftitle{References}
\externalbibliography{yes}
\bibliography{reference}

\begin{thebibliography}{-------}
\providecommand{\natexlab}[1]{#1}

\bibitem[Baggeroer \em{et~al.}(1988)Baggeroer, Kuperman, and
  Schmidt]{baggeroer1988matched}
Baggeroer, A.B.; Kuperman, W.; Schmidt, H.
\newblock Matched field processing: Source localization in correlated noise as
  an optimum parameter estimation problem.
\newblock {\em The Journal of the Acoustical Society of America} {\bf 1988},
  {\em 83},~571--587.

\bibitem[Bogart and Yang(1992)]{bogart1992comparative}
Bogart, C.W.; Yang, T.
\newblock Comparative performance of matched-mode and matched-field
  localization in a range-dependent environment.
\newblock {\em The Journal of the Acoustical Society of America} {\bf 1992},
  {\em 92},~2051--2068.

\bibitem[Baggeroer \em{et~al.}(1993)Baggeroer, Kuperman, and
  Mikhalevsky]{baggeroer1993overview}
Baggeroer, A.B.; Kuperman, W.A.; Mikhalevsky, P.N.
\newblock An overview of matched field methods in ocean acoustics.
\newblock {\em IEEE Journal of Oceanic Engineering} {\bf 1993}, {\em
  18},~401--424.

\bibitem[Mantzel \em{et~al.}(2012)Mantzel, Romberg, and
  Sabra]{mantzel2012compressive}
Mantzel, W.; Romberg, J.; Sabra, K.
\newblock Compressive matched-field processing.
\newblock {\em The Journal of the Acoustical Society of America} {\bf 2012},
  {\em 132},~90--102.

\bibitem[Yang(2014)]{yang2014data}
Yang, T.
\newblock Data-based matched-mode source localization for a moving source.
\newblock {\em The Journal of the Acoustical Society of America} {\bf 2014},
  {\em 135},~1218--1230.

\bibitem[Goodfellow \em{et~al.}(2016)Goodfellow, Bengio, Courville, and
  Bengio]{goodfellow2016deep}
Goodfellow, I.; Bengio, Y.; Courville, A.; Bengio, Y.
\newblock {\em Deep learning}; Vol.~1, MIT press Cambridge,  2016.

\bibitem[Chen \em{et~al.}(2020)Chen, Zhang, and Wang]{chen2020machine}
Chen, R.; Zhang, W.; Wang, X.
\newblock Machine Learning in Tropical Cyclone Forecast Modeling: A Review.
\newblock {\em Atmosphere} {\bf 2020}, {\em 11},~676.

\bibitem[Ghahramani(2003)]{ghahramani2003unsupervised}
Ghahramani, Z.
\newblock Unsupervised learning.
\newblock  Summer School on Machine Learning. Springer,  2003, pp. 72--112.

\bibitem[Lefort \em{et~al.}(2017)Lefort, Real, and
  Dr{\'e}meau]{lefort2017direct}
Lefort, R.; Real, G.; Dr{\'e}meau, A.
\newblock Direct regressions for underwater acoustic source localization in
  fluctuating oceans.
\newblock {\em Applied Acoustics} {\bf 2017}, {\em 116},~303--310.

\bibitem[Niu \em{et~al.}(2017{\natexlab{a}})Niu, Reeves, and
  Gerstoft]{niu2017source}
Niu, H.; Reeves, E.; Gerstoft, P.
\newblock Source localization in an ocean waveguide using supervised machine
  learning.
\newblock {\em The Journal of the Acoustical Society of America} {\bf 2017},
  {\em 142},~1176--1188.

\bibitem[Niu \em{et~al.}(2017{\natexlab{b}})Niu, Ozanich, and
  Gerstoft]{niu2017ship}
Niu, H.; Ozanich, E.; Gerstoft, P.
\newblock Ship localization in Santa Barbara Channel using machine learning
  classifiers.
\newblock {\em The Journal of the Acoustical Society of America} {\bf 2017},
  {\em 142},~EL455--EL460.

\bibitem[Wang and Peng(2018)]{wang2018underwater}
Wang, Y.; Peng, H.
\newblock Underwater acoustic source localization using generalized regression
  neural network.
\newblock {\em The Journal of the Acoustical Society of America} {\bf 2018},
  {\em 143},~2321--2331.

\bibitem[Huang \em{et~al.}(2018)Huang, Xu, Gong, Wang, and
  Yan]{huang2018source}
Huang, Z.; Xu, J.; Gong, Z.; Wang, H.; Yan, Y.
\newblock Source localization using deep neural networks in a shallow water
  environment.
\newblock {\em The Journal of the Acoustical Society of America} {\bf 2018},
  {\em 143},~2922--2932.

\bibitem[Liu \em{et~al.}(2019)Liu, Niu, and Li]{liu2019source}
Liu, Y.N.; Niu, H.Q.; Li, Z.L.
\newblock Source ranging using ensemble convolutional networks in the direct
  zone of deep water.
\newblock {\em Chinese Physics Letters} {\bf 2019}, {\em 36},~044302.

\bibitem[Niu \em{et~al.}(2019)Niu, Gong, Ozanich, Gerstoft, Wang, and
  Li]{niu2019deep}
Niu, H.; Gong, Z.; Ozanich, E.; Gerstoft, P.; Wang, H.; Li, Z.
\newblock Deep-learning source localization using multi-frequency
  magnitude-only data.
\newblock {\em The Journal of the Acoustical Society of America} {\bf 2019},
  {\em 146},~211--222.

\bibitem[Wang \em{et~al.}(2019)Wang, Ni, Su, Hu, Ren, Gerstoft, and
  Ma]{wang2019deep}
Wang, W.; Ni, H.; Su, L.; Hu, T.; Ren, Q.; Gerstoft, P.; Ma, L.
\newblock Deep transfer learning for source ranging: Deep-sea experiment
  results.
\newblock {\em The Journal of the Acoustical Society of America} {\bf 2019},
  {\em 146},~EL317--EL322.

\bibitem[Lin \em{et~al.}(2020)Lin, Zhu, Wu, and Zhang]{lin2020passive}
Lin, Y.; Zhu, M.; Wu, Y.; Zhang, W.
\newblock Passive Source Ranging Using Residual Neural Network With One
  Hydrophone in Shallow Water.
\newblock  2020 IEEE 3rd International Conference on Information Communication
  and Signal Processing (ICICSP). IEEE,  2020, pp. 122--125.

\bibitem[Zhai \em{et~al.}(2019)Zhai, Oliver, Kolesnikov, and
  Beyer]{zhai2019s4l}
Zhai, X.; Oliver, A.; Kolesnikov, A.; Beyer, L.
\newblock S4l: Self-supervised semi-supervised learning.
\newblock  Proceedings of the IEEE international conference on computer vision,
   2019, pp. 1476--1485.

\bibitem[Bianco \em{et~al.}(2020)Bianco, Gannot, and Gerstoft]{bianco2020semi}
Bianco, M.J.; Gannot, S.; Gerstoft, P.
\newblock Semi-supervised source localization with deep generative modeling.
\newblock {\em arXiv preprint arXiv:2005.13163} {\bf 2020}.

\bibitem[Hu \em{et~al.}(2020)Hu, Samarasinghe, Abhayapala, and
  Gannot]{hu2020unsupervised}
Hu, Y.; Samarasinghe, P.N.; Abhayapala, T.D.; Gannot, S.
\newblock Unsupervised Multiple Source Localization Using Relative Harmonic
  Coefficients.
\newblock  ICASSP 2020-2020 IEEE International Conference on Acoustics, Speech
  and Signal Processing (ICASSP). IEEE,  2020, pp. 571--575.

\bibitem[Zeng \em{et~al.}(2013)Zeng, Wang, Zhang, and Cai]{zeng2013feature}
Zeng, X.; Wang, Q.; Zhang, C.; Cai, H.
\newblock Feature selection based on ReliefF and PCA for underwater sound
  classification.
\newblock  Proceedings of 2013 3rd International Conference on Computer Science
  and Network Technology. IEEE,  2013, pp. 442--445.

\bibitem[Ouelha \em{et~al.}(2013)Ouelha, Mesquida, Chaillan, and
  Courmontagne]{ouelha2013extension}
Ouelha, S.; Mesquida, J.R.; Chaillan, F.; Courmontagne, P.
\newblock Extension of maximal marginal diversity based feature selection
  applied to underwater acoustic data.
\newblock  2013 OCEANS-San Diego. IEEE,  2013, pp. 1--5.

\bibitem[Yang \em{et~al.}(2016)Yang, Gan, Chen, Pan, Tang, and
  Li]{yang2016underwater}
Yang, H.; Gan, A.; Chen, H.; Pan, Y.; Tang, J.; Li, J.
\newblock Underwater acoustic target recognition using SVM ensemble via
  weighted sample and feature selection.
\newblock  2016 13th International Bhurban Conference on Applied Sciences and
  Technology (IBCAST). IEEE,  2016, pp. 522--527.

\bibitem[Erkmen and Y{\i}ld{\i}r{\i}m(2008)]{erkmen2008improving}
Erkmen, B.; Y{\i}ld{\i}r{\i}m, T.
\newblock Improving classification performance of sonar targets by applying
  general regression neural network with PCA.
\newblock {\em Expert Systems with Applications} {\bf 2008}, {\em
  35},~472--475.

\bibitem[Jackson(2005)]{jackson2005user}
Jackson, J.E.
\newblock {\em A user's guide to principal components}; Vol. 587, John Wiley \&
  Sons,  2005.

\bibitem[Westad \em{et~al.}(2003)Westad, Hersletha, Lea, and
  Martens]{westad2003variable}
Westad, F.; Hersletha, M.; Lea, P.; Martens, H.
\newblock Variable selection in PCA in sensory descriptive and consumer data.
\newblock {\em Food Quality and Preference} {\bf 2003}, {\em 14},~463--472.

\bibitem[Murray and Ensberg(1996)]{murray1996swellex}
Murray, J.; Ensberg, D.
\newblock The swellex-96 experiment.
\newblock {\em URL: http://www. mpl. ucsd. edu/swellex96} {\bf 1996}.

\bibitem[Esbensen \em{et~al.}(2002)Esbensen, Guyot, Westad, and
  Houmoller]{esbensen2002multivariate}
Esbensen, K.H.; Guyot, D.; Westad, F.; Houmoller, L.P.
\newblock {\em Multivariate data analysis: in practice: an introduction to
  multivariate data analysis and experimental design}; Multivariate Data
  Analysis,  2002.

\bibitem[H{\o}y \em{et~al.}(2002)H{\o}y, Westad, and Martens]{hoy2002combining}
H{\o}y, M.; Westad, F.; Martens, H.
\newblock Combining bilinear modelling and ridge regression.
\newblock {\em Journal of Chemometrics: A Journal of the Chemometrics Society}
  {\bf 2002}, {\em 16},~313--318.

\bibitem[Hinton and Zemel(1994)]{hinton1994autoencoders}
Hinton, G.E.; Zemel, R.S.
\newblock Autoencoders, minimum description length and Helmholtz free energy.
\newblock  Advances in neural information processing systems,  1994, pp. 3--10.

\bibitem[Chen \em{et~al.}(2017)Chen, Shi, Zhang, Wu, and Guizani]{chen2017deep}
Chen, M.; Shi, X.; Zhang, Y.; Wu, D.; Guizani, M.
\newblock Deep features learning for medical image analysis with convolutional
  autoencoder neural network.
\newblock {\em IEEE Transactions on Big Data} {\bf 2017}.

\end{thebibliography}

%% for journal Sci
%\reviewreports{\\
%Reviewer 1 comments and authors’ response\\
%Reviewer 2 comments and authors’ response\\
%Reviewer 3 comments and authors’ response
%}

%%%%%%%%%%%%%%%%%%%%%%%%%%%%%%%%%%%%%%%%%%
\end{document}